\documentstyle[epsfig,amssymb,floatflt,float,here]{mn2e}
\newif\ifAMStwofonts

\newcommand{\sm}{\, {\rm M}_{\odot}}

\newcommand{\kms}{km~s$^{-1}$}
\newcommand{\D}{\displaystyle}

\title[The phase-space structure of cold dark-matter halos]{The
phase-space structure of cold dark-matter halos:\\ Insights into the
Galactic halo} \author[A. Helmi, S.D.M. White and V. Springel] {Amina
Helmi\thanks{Present address: Sterrenkundig Institut, Universiteit
Utrecht, P.O.Box 80000, 3508 TA Utrecht, The Netherlands. E-mail:
a.helmi@phys.uu.nl}, Simon
D.M. White\thanks{E-mail:swhite@mpa-garching.mpg.de} and Volker
Springel\thanks{E-mail:volker@mpa-garching.mpg.de} \\ Max-Planck-Institut
f\"ur Astrophysik, Karl-Schwarzschild-Str.  1, 85740 Garching bei
M\"unchen, Germany} \date{Accepted ...  Received ...; in original form
...}

\pagerange{\pageref{firstpage}--\pageref{lastpage}}
\pubyear{}

\begin{document}

\maketitle

\label{firstpage}

\begin{abstract}
We study the formation of the Milky Way's halo in a $\Lambda$CDM
cosmology by scaling down a high resolution simulation of the
formation of a cluster of galaxies. We determine how much phase-space
substructure is left over from the objects that merge to build up the
present galaxy. We study the debris streams originating from such
objects and find that their evolution can be explained simply in terms
of the conservation of phase-space density.  Analysing the mass growth
history of our halo we find that its inner regions have been in place
for more than 10 Gyr, but that the growth of the halo as a whole is
more gradual, in agreement with other high resolution simulations of
dark-matter halos. Recent accretion contributes to the inner 10 kpc of
the halo only at the 10$^{-4}$ level. Finally we determine the number
of dark-matter streams as a function of distance from the centre of
the halo. In the equivalent of the ``Solar vicinity'', we find that
the dark-matter is smoothly distributed in space, and that the
velocity ellipsoid is formed by hundreds of thousands of streams, most
of which have velocity dispersions of the order of 1 \kms~ or less.
\end{abstract}


\section{Introduction}

Over the last twenty years, the hierarchical paradigm has emerged as
the standard model to describe the formation of structure in the
Universe. As embodied in the current ``concordance" $\Lambda$CDM model
it appears to be consistent with a very wide range of cosmological
data ranging from fluctuations in the Cosmic Microwave Background
through the structure of Ly$\alpha$ forest absorption in QSO spectra
and the gravitational shear induced by dark-matter structures to the
observed large scale structure in the galaxy distribution. An
important characteristic of such models is that they are based on a
set of well-defined and testable assumptions. This renders possible
the detailed modelling of the formation and evolution of galactic
systems, and a later comparison to observations of the properties of
these systems as a function of environment or redshift (e.g. Diaferio
et al. 2001; Benson et al. 2001; Somerville, Primack \& Faber 2001).

 It is also possible to test the hierarchical paradigm on our Galaxy
(e.g. Hern\'andez, Avila-Reese \& Firmani 2001).  Several groups
(Moore et al. 1999; Klypin et al. 1999; Klypin, Zhao \& Somerville
2002) have focused on the properties of dark halos, hoping to
constrain the nature of dark-matter. These groups performed high
resolution simulations of a galactic size halo in CDM
cosmologies. They confirmed earlier analytic claims (Kauffmann, White
\& Guiderdoni 1993) that the predicted number of satellites exceeds
the known population in the Local Group by a factor of ten.  Some
attempts have been made to account for the disagreement, by changing
the nature of the dark-matter (Spergel \& Steinhardt 2000; Bode,
Ostriker \& Turok 2001), by modifying the initial power spectrum of
density fluctuations (Kamionkowski \& Liddle 2000) or by taking into
account the effects of a reionising background which may inhibit star
formation in the smallest mass halos (Kauffmann et al. 1993; Bullock,
Kravtsov \& Weinberg 2001; Benson et al. 2002). The recent results by
Kleyna et al. (2002) on the mass distribution in the Draco dSph (see
also Mateo 1997 and Lokas 2001 for a similar study on Fornax) favour
an astrophysical explanation since the actual circular velocities of
the other satellite galaxies of the Milky Way are in fact several tens
of \kms~ larger than previously thought, and agree with those expected
for the most massive substructures in a $\Lambda$CDM universe (Stoehr
et al. 2002).

Broadly speaking, the hierarchical paradigm predicts that the Milky
Way formed through mergers of smaller systems (White \& Rees 1978).
These systems would contribute to the dark halo, the spheroid (the
bulge and the stellar halo) and to the Galactic gas reservoir. It may
be very difficult to determine the relative gas contribution of these
progenitor objects to the present Galaxy, since gas ``easily forgets''
its site of origin. However for collisionless stars and dark-matter
the situation can be quite different. If the dynamical mixing
timescales are sufficiently long (i.e. longer than the age of the
Universe) it may be possible to ``break-up'' the Galactic spheroid
(stars and may be even dark matter particles) into coherent structures
in phase-space directly related to the systems that merged to form the
Milky Way we observe today.

A first attempt at determining whether the merging history of the
Milky Way may be imprinted in the phase-space structure of {\it
nearby} halo stars, and thus be recovered, was made by Helmi \& White
(1999; hereafter HW). They studied the infall of satellites onto a
fixed Galactic potential, and the evolution of the debris in
phase-space. They found that after 10 Gyr stars having a common origin
are distributed smoothly in space, but appear very clumped in velocity
space, where they define streams with very small velocity
dispersions. The expected number of such streams scales with the
initial size $r$, velocity dispersion $\sigma$ and orbital period $P$,
of the disrupted object:
\begin{equation}
N_{\rm stream} \sim 10 \D\left(\frac{r}{1~{\rm kpc}}\right)^2
\left(\frac{\sigma}{15~{\mbox \kms}}\right) \left(\frac{P}{0.23~{\rm
Gyr}}\right)^{-3}.
\end{equation}
The total number of stars associated with the object is $N_* \propto
r\sigma^2$ (from the virial theorem) while the volume $V$ over which
they are spread scales with the cube of the size of the orbit, and so
approximately as $P^3$. Hence the number of stars per stream in the
Solar neighbourhood scales as $N_*/VN_{\rm stream} \sim \sigma /r$;
objects with large initial velocity dispersion and small initial size
should produce the most easily detectable streams with little
dependence on initial period.  Such arguments suggest that the Solar
neighbourhood velocity ellipsoid is composed of $300-500$
kinematically coherent structures which originated in past merger and
accretion events. A pair of halo streams that can perhaps be directly
linked to a disrupted satellite were detected in the Solar
neighbourhood by Helmi et al.(1999).  The progenitor of these two
streams was probably similar to the dwarf galaxy Fornax. 
Substructure in the outer halo also appears to be ubiquitous, and has
been found by several surveys over the last few years (e.g. Ivezic et
al.  (2000) and Yanny et al. (2000) for the SDSS; Dohm-Palmer et
al. (2001) for the SPS; Vivas et al. (2001) for QUEST).  Most of these
recently discovered structures can be associated to just one of Milky
Way's satellites: the Sagittarius dwarf galaxy which is in the process
of being completely disrupted (Ibata et al. 2001;
Mart\'{\i}nez-Delgado et al. 2001; Helmi \& White 2001).

A weak point of the HW analysis and of similar studies (e.g. Johnston,
Hernquist \& Bolte 1996; Johnston 1998), is the assumption of a fixed,
smooth potential onto which galaxies are accreted. In hierarchical
clustering, galaxy potentials are constantly changing, and can vary
very violently during mergers. Large numbers of clumps orbit the
centre of even a ``virialised" halo. These clumps may have substantial
effects on the structure of debris streams (e.g Johnston, Spergel \&
Haydn 2002; Ibata et al. 2002; Mayer et al. 2002).

The main goal of the present paper is to understand the phase-space
structure of cold dark-matter halos. In particular, we want to study
the evolution of satellite debris, and to quantify the expected amount
of substructure. We also want to determine to what extent previous
results are valid in the truly hierarchical regime of the build--up of
a galaxy.  We tackle these problems by scaling down to galactic
size a high-resolution simulation of the formation of a cluster in a
$\Lambda$CDM cosmology (Springel et al. 2001). 

The paper is organised as follows. In Sec.2 we describe the
simulations, in Sec.3 we follow the evolution in phase-space of debris
streams, and compare to the analytic model of HW in Section 3.2. Sec.4
describes the mass-growth history of the simulated dark-matter halo,
and in Section 5 we determine the number of streams and their internal
properties as function of distance from the dark-matter halo
centre. We leave the summary and discussion of our results for Section
6.
\section{Methodology}

\subsection{The simulations}

The simulation we analyse here was carried out using a parallel
tree-code (Springel, Yoshida \& White 2001) on the Cray T3E at the
Garching Computing Centre of the Max Planck Society. Its initial
conditions were generated by zooming in and re-creating with higher
resolution a particular galaxy cluster and its surroundings formed within a
cosmological simulation (as in Tormen, Bouchet \& White 1997). The
original parent $\Lambda$CDM cosmological simulation (from Kauffmann et
al. 1999) had parameters $\Omega_0 = 0.3$, $\Omega_\Lambda = 0.7$, $h=
0.7$ and $\sigma_8 = 0.9$. The cluster selected for re-simulation was
the second most massive cluster in this simulation, having a virial
mass of $8.4 \times 10^{14} h^{-1}\sm$. The particles that end up in
the final cluster of the cosmological simulation and in its immediate
surroundings (defined by a comoving sphere of $70 h^{-1}$ Mpc radius)
were traced back to their Lagrangian region in the initial conditions
for re-simulation. The initial mass distribution between $21$ and $70
h^{-1}$ Mpc was represented by $3 \times 10^6$ particles.  In the
inner region, where the original simulation had $2.2 \times 10^5$
particles, new initial conditions were created using $6.6 \times 10^7$
particles.  Small scale power was added in accordance with the better
k-space sampling allowed by the larger number of particles. The
original force softening was also decreased to obtain better spatial
resolution.  The new simulation was run from high redshift until
$z=0$, and was analysed in considerable detail in Springel et
al. (2001). In this high resolution simulation there are about 20
million particles within the virialised region of the cluster halo.

\subsection{Scaling to a Milky Way halo}

\begin{figure}
\psfig{figure=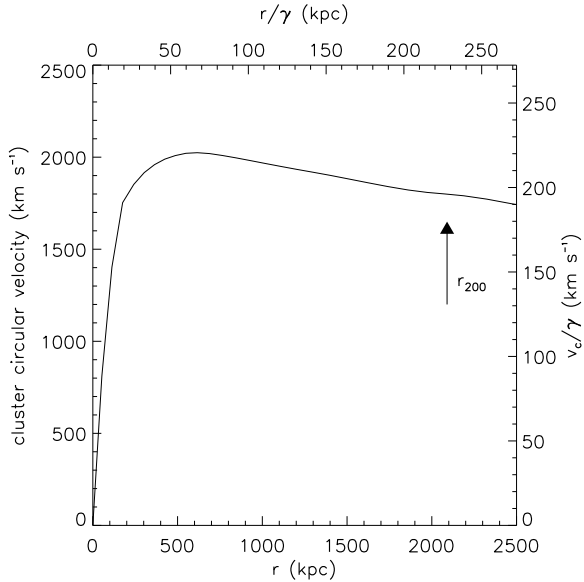,height=8cm}
\caption{The circular velocity $V_c(r) = \sqrt{GM(r)/r}$ of the
cluster for our simulation S4. Axes show radii and circular velocities
both before and after scaling down to Milky Way size.}
\label{lcdm_fig:circ_veloc}
\end{figure}

In Figure~\ref{lcdm_fig:circ_veloc} we show the circular velocity
profile of the cluster at $z=0$.  We determine the cluster centre by
successively refining a mesh located on the cluster, and determining
the cell containing the largest number of particles. This process is
repeated until the largest number of particles in a given (now small
size) cell is sufficiently small to determine by simple counts which
particle has the largest number of neighbours.  Its position
corresponds to the point of maximum density and it is defined as the
cluster centre.  This determination is robust against changes in the
mesh shape and size, and we estimate the error in the final position
of the centre of the cluster to be of the order of 0.7 kpc, comparable
to the scale of the gravitational softening used in the simulation.
The circular velocity is then determined after spherically averaging
the mass distribution around the centre of the cluster and is derived
from $V_c(r) = \sqrt{GM(r)/r}$.
By fitting a Navarro, Frenk \& White (NFW; 1996) profile, we find that
the concentration of the cluster halo is $c_{\rm NFW} = 7.3$, the
scale radius $r_s = 285$~kpc and the virial radius $r_{200} =
2090$~kpc. 

We can scale the cluster to a ``Milky Way'' halo, by requiring that
its maximum circular velocity be equal to 220 \kms. The required
scaling factor $\gamma$ is given by 
\begin{equation}
\label{eq:scale}
\gamma = \frac{V_{c}^{\rm cl}}{V_{c}^{\rm MW}} = 
	\frac{r_{\rm 200}^{\rm cl}}{r_{\rm 200}^{\rm MW}} \sim 9.18. 
\end{equation}
 With this scaling, the virial radius of our simulated Milky Way
dark-matter halo is 228~kpc.  Its virial mass is $1.28 \times 10^{12}
\sm$, the mass of an individual particle is $8.66 \times 10^4 \sm$,
and the gravitational force in the final object has an equivalent
Plummer softening of 0.11 kpc.

Our argument that with the simple scaling of Eq.(\ref{eq:scale}) this
simulation is a fair representation of the growth of the Milky Way
relies on both theoretical and numerical results (Lacey \& Cole 1993;
Moore et al. 1999). Numerical simulations by Moore et al. (1999) have
shown that galaxy and cluster halos have similar properties in terms
of final structure (density profile), number of satellites, etc.,
despite typically assembling at systematically different
redshifts. Jing \& Suto (2000) have also performed high resolution
simulations of several galaxy, group and cluster size halos. Although
they find that galaxy halos have a steeper inner profile than larger
mass halos, they also find that the scatter in the properties of
objects belonging to the same class is as large as the systematic
differences between classes.  We expect therefore, that our scaled
simulation will represent reasonably well the formation process of a
galactic halo, except that its assembly occurs at later times than the
majority of such halos. From now on we will use this scaling in the
paper, and assume, unless otherwise stated, that the simulation
represents a galaxy halo.

\section{The phase-space evolution of debris}
\label{lcdm_sec:phase-space}

Here we study the phase-space evolution of debris from the disrupted
halos that end up forming the ``Galactic" dark halo at the present
time. With this goal in mind, we proceed by identifying halos at
high-redshift which are directly accreted onto the main progenitor of
the ``Milky Way's" halo.

We identify halos at each output time using a Friends-of-Friends (FOF)
algorithm, which links particles separated by less than 20\% of the
mean interparticle separation.  In this way we can construct a 
detailed merger history of the galaxy. As we step back in redshift we
identify at each output the most massive halo which is part of the
galaxy's main progenitor at the subsequent time.  We say that a halo
identified at redshift $z$ will be directly accreted onto the main
progenitor at $z'$ (the redshift of the next simulation output) if at
least half of its particles {\it and} the most-bound particle have
become part of the main progenitor at $z'$. Here the most-bound
particle refers to the particle with the minimum potential energy in
the (to be) accreted halo.

The centre of mass position ${\bf p}_{CM}$ and velocity ${\bf v}_{CM}$
of the main progenitor are computed as follows. We first determine the
location of the most bound particle ${\bf p}_{\rm mb}$. We define
concentric spheres of successively smaller radii (down to 10 kpc
radius) around this particle, and compute the centre of mass ${\bf
p'}_{CM}$ from the particles within these spheres. We stop this
iterative procedure when $| {\bf p}_{\rm mb} - {\bf p'}_{CM}| < \epsilon
$, where $\epsilon = 0.3$ kpc. This is then defined as the location of
centre of mass of the main progenitor. The velocity of the centre of
mass is then ${\bf v'}_{CM}$, measured by the velocities of the
particles within the largest sphere for which the above condition is
satisfied.
\begin{figure}
\psfig{figure=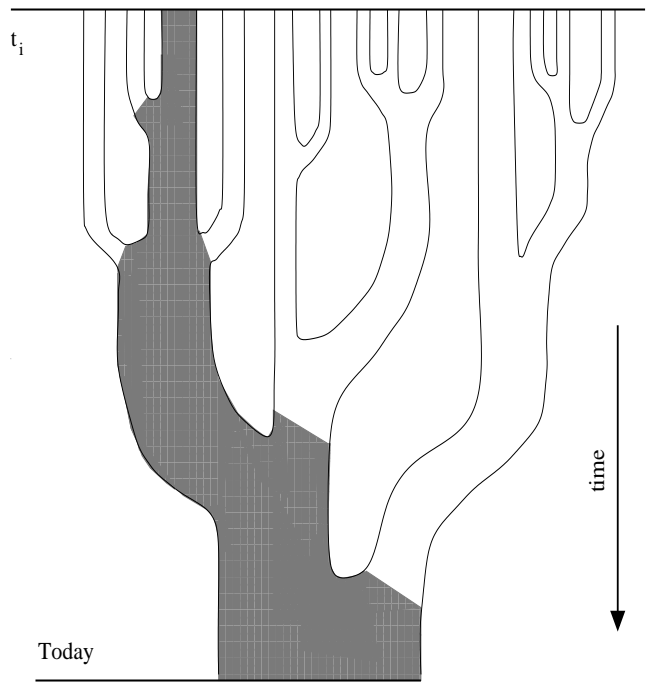,height=9.cm,width=8.5cm}
\caption{A schematic representation of the merger history of a galaxy.
The dark trunk at any given time represents the galaxy main
progenitor, and the branches linked to this trunk correspond to the
halos that merge directly with the galaxy (based on Lacey \& Cole,
1993). }
\label{lcdm:fig_mergerTree}
\end{figure}

This is like following the (thickest) trunk of the merger tree, which
would correspond to the galaxy's main progenitor, and studying what
happens to halos which join from other tree branches as time
progresses. The idea is illustrated in
Figure~\ref{lcdm:fig_mergerTree}.

\begin{figure}
\psfig{figure=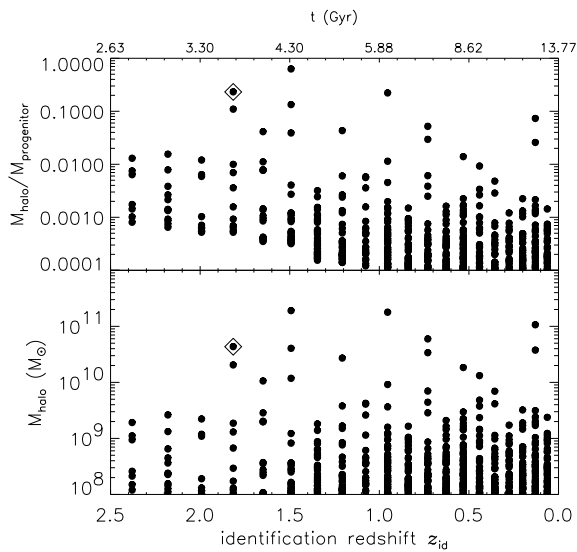,width=8.5cm}
\caption{The top panel shows the mass ratio of the halos directly
accreted onto the main progenitor as a function of their accretion
time. Note that in some cases this mass ratio approaches one, meaning
that the galaxy is experiencing a major merger. In the bottom panel
we plot the actual masses of these halos. The diamonds are here used
to denote a halo which we discuss in detail in Sections 3.1 and 3.2.}
\label{lcdm:fig_halosOnCluster}
\end{figure}

For our high resolution simulation we have identified 752 halos with
at least 1000 particles (which corresponds to a minimum mass of
$8.66\times 10^7 \sm$), which fall onto the main progenitor between
redshift $z = 2.4$ and the present day. The accreted halos
have a large spread in mass as shown in the bottom panel of
Figure~\ref{lcdm:fig_halosOnCluster}.  From the top panel of this
Figure we note that in some cases the satellite-to-primary mass ratio
is close to unity, corresponding to a major merger. Such mergers,
although few in number, contribute a substantial fraction of the total
mass growth.

\subsection{Spatial and kinematic evolution of halo debris}

We now study in detail how the debris streams from one of the above
752 halos evolve in time. We follow a halo that merges with the main
progenitor at redshift 1.8, and has an initial mass of $4.3 \times
10^{10} \sm$. This halo is denoted with a diamond in
Figure~\ref{lcdm:fig_halosOnCluster}. Its mass at the time of
accretion was about 25\% that of the main progenitor. This satellite
decays through dynamical friction to the centre of the main progenitor
where it is fully disrupted. The mass stripped off at earlier times
mainly populates the outer regions of the final object, while that
lost at late times ends up closer to its centre.

\subsubsection{Evolution of debris in the outer galaxy}

\begin{figure*}
\psfig{figure=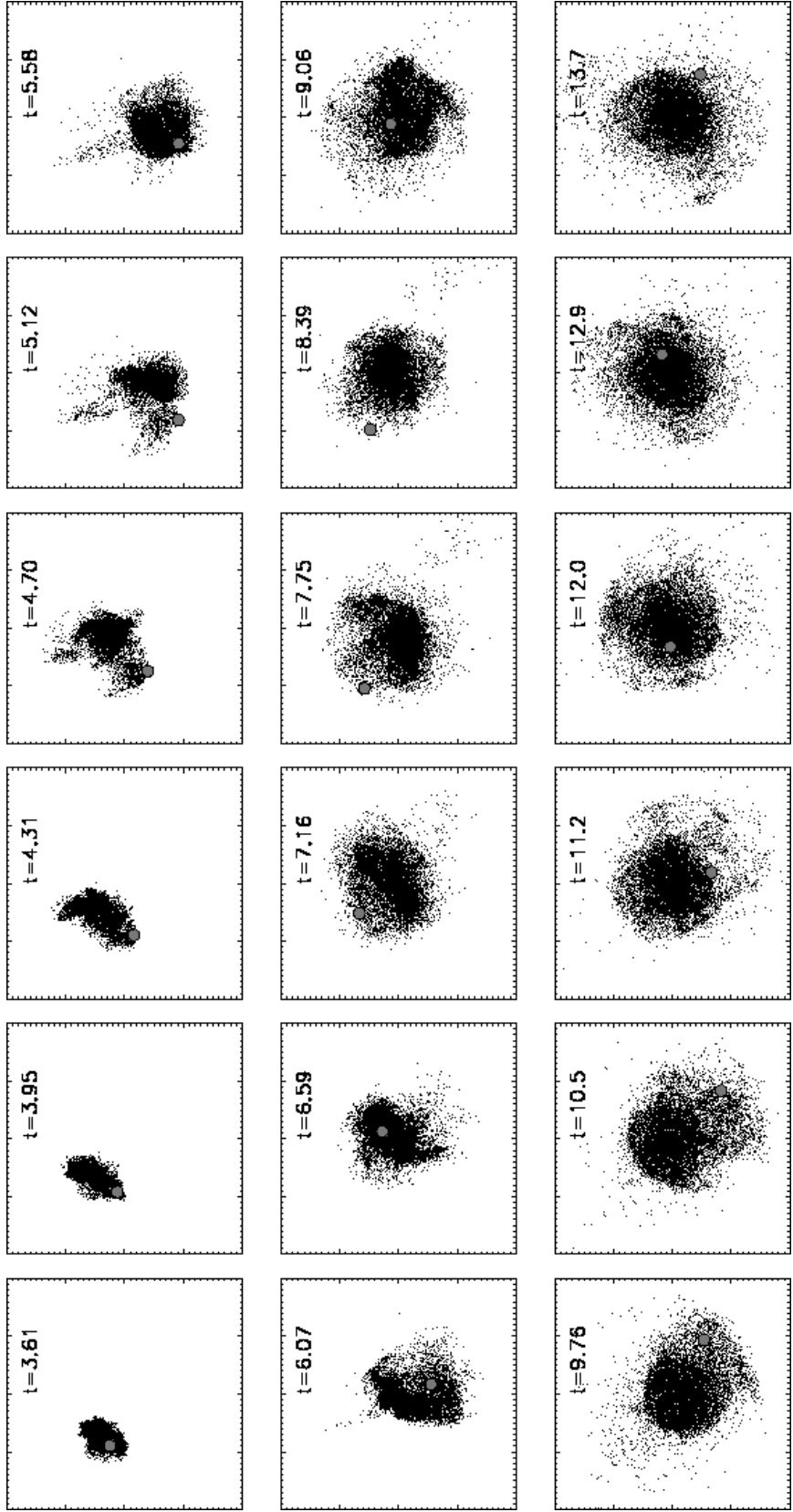,height=22cm,angle=180}
\caption{Evolution of the spatial distribution of particles from a
halo of $4.3 \times 10^{10} \sm$ that merged at $z = 1.8$ ($t=3.61$).
We plot here the $x$ vs $z$ positions of the 10\% of the particles
that were lost from this halo at this redshift. The box is 400 kpc (in
physical units) on a side, and at each output it is centred on the
main progenitor's centre of mass. The grey solid circle denotes the
location of the reference particle which belongs to a stream orbiting
the outer galaxy. }
\label{fig:pos_halo02}
\end{figure*}

We identify material from the accreted halo which is part of a tidal
stream in the outer galaxy at the present time.  We select a reference
particle in this structure, which can be traced back to $z=1.8$, and
then followed forwards in time.  In Figure \ref{fig:pos_halo02} we
show a time sequence of the spatial distribution of particles lost at
$z = 1.8$. We say that a particle has been lost by its progenitor halo
if its binding energy has become positive. The binding energy
$\epsilon$ of a particle located at ${\bf r}$ and with velocity ${\bf
v}$ with respect to the centre of mass of the satellite\footnote{The
centre of mass is here defined by the 0.1\% most bound particles of
the system, which in this case corresponds to 500 particles.} is
defined as $\epsilon = \Phi({\bf r}) + 0.5 |{\bf v}|^2$, where
$\Phi({\bf r})$ is the potential energy at ${\bf r}$ due to all
particles in the satellite.  From Fig.~\ref{fig:pos_halo02} we see
that the initial distribution of particles is relatively compact, and
that as time passes by the material is strung out in a characteristic
stream-like structure (e.g at $t=4.31$ Gyr) over several tens to
hundreds of kiloparsecs. At the final time the material appears to be
more smoothly distributed over the whole box, which is 400 kpc (in
physical units) on a side.  Our simulations are not well-suited to
address the effects of other dark-matter lumps on tidal tails (cf
Johnston et al. 2001), which could be related to some of the transient
structures observed in this Figure. Our main limitation is the large
time interval between stored outputs which prevents us from
determining the effect of close encounters and their relation to the
features seen in Figure~\ref{fig:pos_halo02}.

In many of the snapshots tight small substructures can be
observed. These objects were already present as subhalos within the
satellite halo before it was accreted by the galaxy. After accretion,
they were released from their parent satellite, becoming
subhalos of the main galaxy.  It is worth mentioning that these
subhalos only constitute a small fraction of the debris material lost
by the satellite (less than 10\%, e.g. Ghigna et al. 2000; Springel et
al. 2001). 

\begin{figure}
\psfig{figure=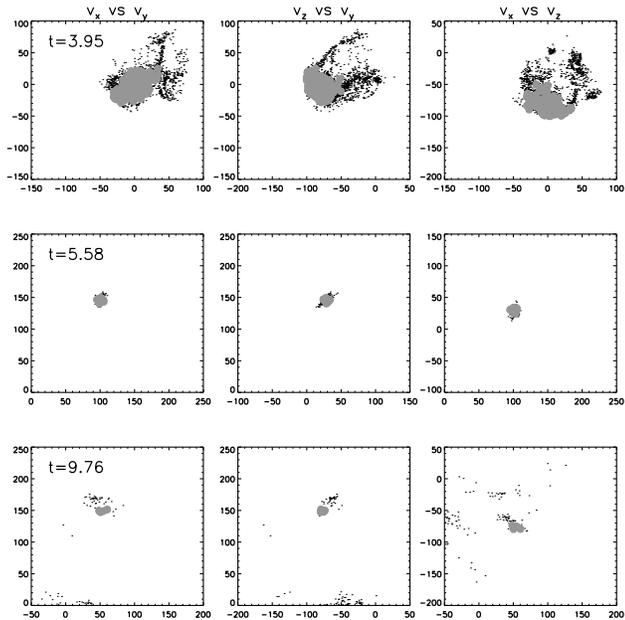,width=8.5cm}
\caption{Evolution of the velocity distribution of neighbours of the
reference particle (shown as a solid grey circle in Figure
\ref{fig:pos_halo02}) of the satellite halo of the previous figure.
Dots correspond to particles that fall inside a cube of side 20 kpc
around the reference particle at the given time. The filled circles
are particles that fall within a box of side 10 kpc, and were
neighbours of the reference particle in all previous outputs. These
particles define the stream in which the reference particle is
located. Notice the decrease in the density and velocity dispersion of
the filled circles.}
\label{fig:vel_halo02}
\end{figure}

In Figure \ref{fig:vel_halo02} we show the velocities of debris
particles that are relatively close (inside a cube of side 20 kpc)
to the reference particle, at three different times. For comparison,
note that the virial radius of the satellite at the time of infall was
approximately 37 kpc. The solid grey circles correspond to particles
which
\begin{enumerate}
\item were always neighbours of the reference particle: to be in this
set, particles need to have $x$, $y$ and $z$ coordinates within 10 kpc
of those of the reference particle in all previous outputs;

\item are within 5 kpc of the reference particle in each coordinate in
the current output.
\end{enumerate}
  We note that the initial velocity distribution of the halo is broad,
and relatively clumpy. As discussed before, this clumpiness reflects
the internal structure of the object that fell in.  As time goes by,
the motions of the neighbours of the reference particle become more
similar, and the velocity-box is empty except for velocities close to
that of the reference particle (middle row). At later times (bottom
row), other moving groups are visible, showing that the system has now
produced multiple intersecting streams even in the outer galaxy.
\begin{figure}
\psfig{figure=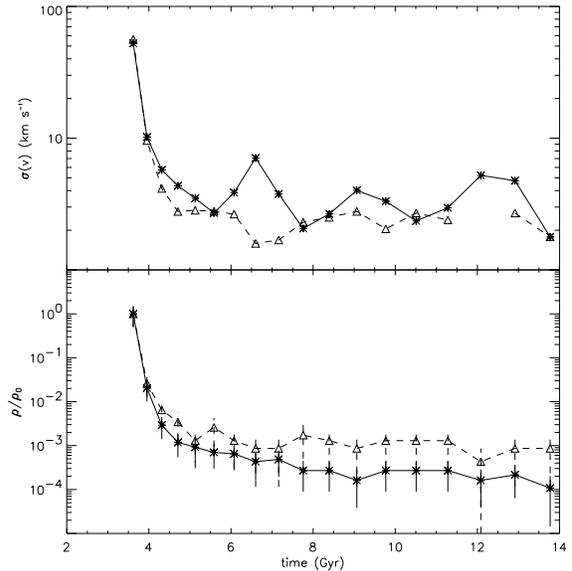,width=8.5cm}
\caption{The top panel shows the evolution of the velocity dispersion
in the neighbourhood of the reference particle of the same satellite
halo as in Figures \ref{fig:pos_halo02} and \ref{fig:vel_halo02}. The
solid line corresponds to the 1--d velocity dispersion computed from
the particles shown as grey solid circles in
Fig.~\ref{fig:vel_halo02}. The dashed curve is for a subset which are
even closer to the reference particle, within a box of side 5 kpc. The
differences can be attributed to gradients along the orbit, which are
particularly important at late times. The bottom panel shows the
evolution of the density of the stream defined by the same set of
particles used in the top panel. The long term behaviour is probably
not physical, especially in the case of the smaller box, but is
dominated by Poisson noise in the number of particles counted.}
\label{fig:vel_ellip02}
\end{figure}

The evolution of the velocity dispersion $\sigma$ in the neighbourhood
of the reference particle is shown in the top panel of
Figure~\ref{fig:vel_ellip02}. Here $\sigma$ is defined as
$\sqrt{\sigma_x^2 + \sigma_y^2 + \sigma_z^2}/\sqrt{3}$, and is
therefore independent of the choice of the coordinate system. It is
measured for the set of particles shown as grey solid circles in
Figure~\ref{fig:vel_halo02}, that is for particles that satisfy both
conditions (i) and (ii) for being neighbours of the reference
particle. We also measure $\sigma$ for a subset of even closer
neighbouring particles (whose $x$, $y$, and $z$ coordinates in all
outputs before the one under study were within 5 kpc of those of the
reference particle, and in the current output are within 2.5 kpc). The
different values of the velocity dispersion in the stream obtained for
these two cases are due to velocity gradients, which can be as large
as the measured dispersions themselves. Note as well the decrease by
roughly a factor 10 in the velocity dispersions in only 2 Gyr.

The stream's density is shown in the bottom panel of the same
figure. It is measured at each output by the number of neighbours in
the stream in boxes of a given size (either of 5 or 10 kpc on a side)
around the reference particle. For the larger volume, this number
evolves from being slightly larger than five thousand to only thirteen
after less than 2 Gyr of evolution. This implies that after this time,
the number counts are dominated by Poisson noise.

\subsubsection{Evolution of debris closer to the centre of the galaxy}

To follow the evolution of streams closer to the centre of the galaxy
we focus on material lost a few Gyr after the satellite halo shown as
a diamond in Fig.\ref{lcdm:fig_halosOnCluster} was accreted.  We here
focus on debris lost from this satellite halo at $t=7.16$ Gyr, or 3.5
Gyr after infall. This is deposited at an intermediate distance from
the centre of the galaxy.  We decided against tracking material lost
even later because it turns out to be very difficult to follow streams
from such material with our numerical resolution and number of stored
outputs. The material lost at later times mixes on shorter
timescales. Streams more rapidly reach very small densities, and we
are unable to determine their properties reliably.

\begin{figure*}
\psfig{figure=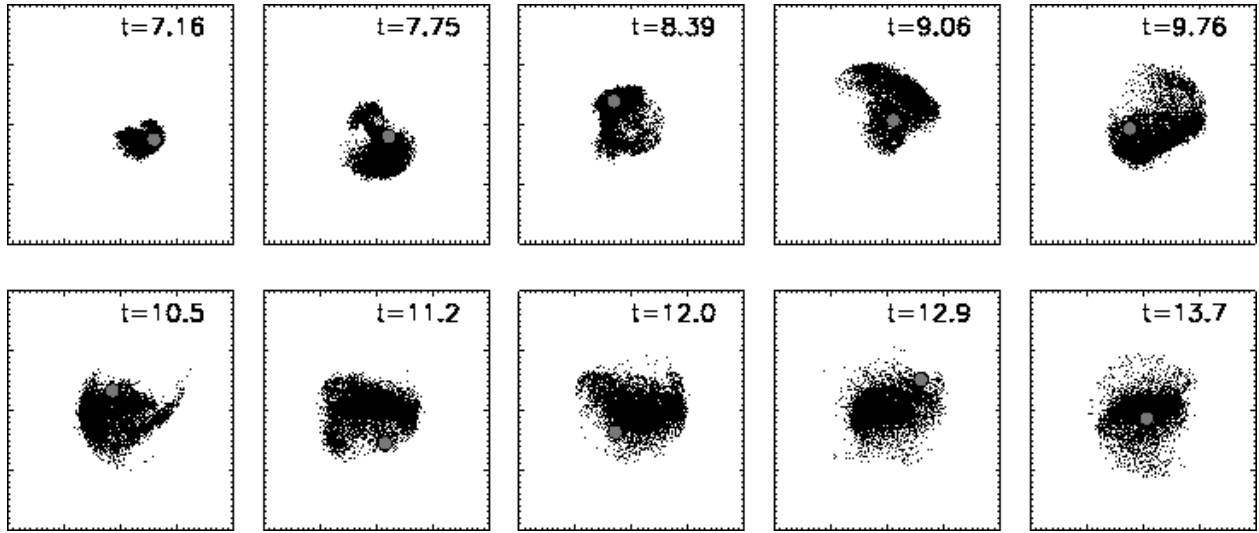,height=7cm}
\caption{Evolution of the spatial distribution of particles from a
halo of $4.3 \times 10^{10} \sm$ that merged with the galaxy at $z =
1.8$.  We plot here only 25\% of the particles lost from this halo 3.5
Gyr after infall (6.6 Gyr ago).  As in Figure 4, the box is 400 kpc on
a side in physical units, and its origin coincides with the position
of the main progenitor's centre of mass at all times. We only show the
$x$ (horizontal axis) vs $y$ (vertical axis) projection. The material
lost at this time is deposited within about 100 kpc from the galaxy's
centre, much closer than that lost at earlier times (for comparison
see Fig.\ref{fig:pos_halo02}). We note that there is a fairly sharp
cutoff in the density at that radius. This corresponds roughly to the
apocentre of the orbit of the satellite at the time these particles
were released.  The grey solid circle shows the location of the
reference particle which belongs to a stream orbiting the
``intermediate" galaxy at the present time.}
\label{fig:pos_halo02b}
\end{figure*}
\begin{figure}
\psfig{figure=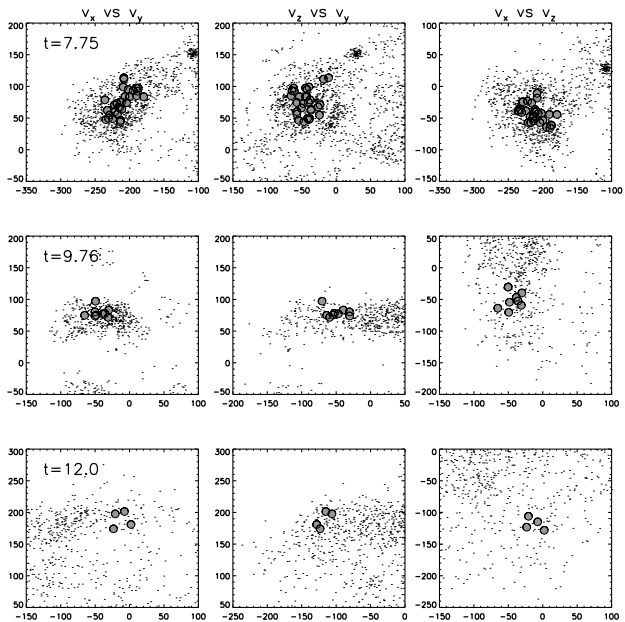,width=8.5cm}
\caption{Evolution of the velocity distribution of neighbours to the
reference particle of the same halo as before, where the reference
particle orbits the intermediate regions of the galaxy at the present
time.  Dots correspond to particles that fall within a box of side 20
kpc from the reference particle in the output shown.  The solid grey
circles correspond to particles that fall within a box of side 10 kpc,
and were also neighbours of the reference particle at all previous
times. They define the stream to which the reference particle
belongs.}
\label{fig:vel_halo02b}
\end{figure}
\begin{figure}
\psfig{figure=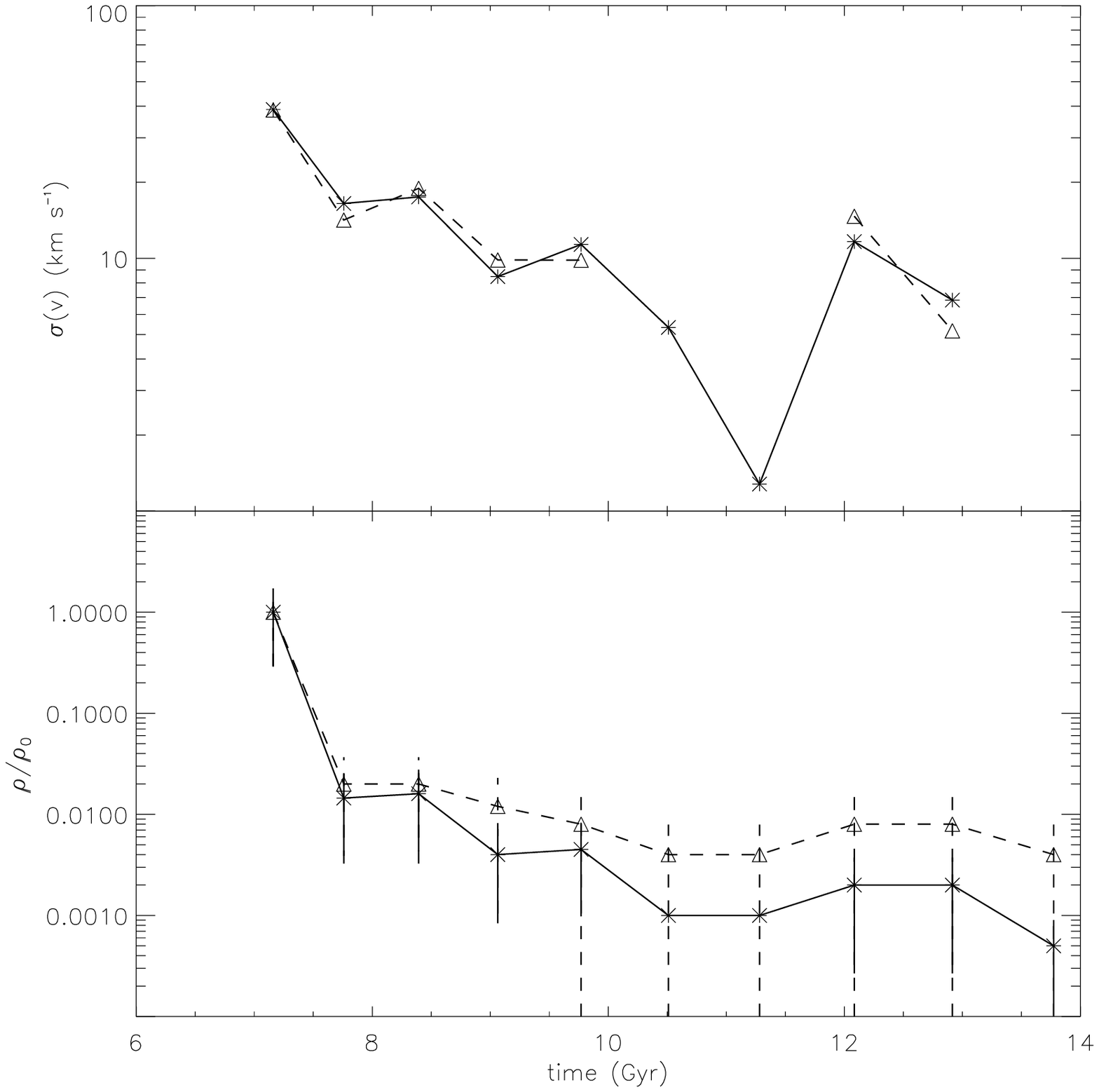,width=8.5cm}
\caption{The top panel shows the evolution of the 1-d velocity
dispersion in the neighbourhood of the reference particle of the same
halo as in the previous figure, and for a stream orbiting the
``intermediate" regions of the galaxy at the present time. The solid
curve corresponds to the particles shown as grey solid dots in
Fig.~\ref{fig:vel_halo02b}, whereas the dashed curve corresponds to a
subset of these located in a volume of side 5 kpc around the
reference particle. The difference between the dashed and solid curves
is due to velocity gradients which are dominant in these relatively
large boxes. The bottom panel shows the evolution of the density of
the stream of the top panel. The error bars are estimates of the noise
based on Poissonian counts. Note that the values of the velocity
dispersion after 10.5 Gyr are determined from the velocities of just
two or three particles.}
\label{fig:vel_ellip02b}
\end{figure}

Figure \ref{fig:pos_halo02b} shows the evolution of the spatial
distribution of the particles lost at $t=7.16$ Gyr. Note that the
particles are more smoothly distributed than in the case of the
outer halo material shown in Figure~\ref{fig:pos_halo02}. This is
because the inner regions of (satellite) halos are in general much
less lumpy than the outskirts due to the shorter dynamical timescales.

As in the previous section we choose a reference particle (amongst all
those lost at $t=7.16$ Gyr) that orbits through the intermediate
regions of the galaxy.  Proceeding as before, we follow the evolution
of the velocities of particles located close to the reference particle
at all times.  Figure~\ref{fig:vel_halo02b} shows the velocity
distribution in the neighbourhood of the reference particle. By
comparing to the analogue for the outer halo stream
(Fig.~\ref{fig:vel_halo02}) we note that, not just the spatial
distribution, but also the velocity distribution is much smoother
initially. This is also true at late times, and very rapidly a regime
is reached where multiple streams can be observed.
Figure~\ref{fig:vel_ellip02b} corresponds to the evolution of the
velocity dispersion and density of a stream.  Again we are not
able to follow the evolution of the velocity dispersion of the stream
at late times, because we run out of neighbouring particles quite
rapidly, particularly for the smaller box.  After $t = 10.5$ Gyr, or 2
Gyr of evolution, the number of neighbours is dominated by Poisson
noise. After this time, the values of the 1-d velocity dispersion,
when measured, are based on the velocities of only two or three
particles.

The evolution of the intermediate and outer halo streams is
characteristic of all streams originating in directly accreted
halos. Naturally the properties of streams and their location in the
galaxy halo will depend on their progenitor, in particular on its
initial mass. In this respect, streams originating in smaller halos
are narrower, more clearly defined, and, typically, they phase-mix on
longer timescales.

\subsection{Mixing in phase-space: Comparison to analytic models of
stream evolution}
\label{sec:mixing}

In the previous section we found a very rapid decline in the density
and velocity dispersion of streams, especially in the first few Gyr of
evolution. After this initial period, the number of particles in a
stream is so small in our simulation that it is dominated by
Poisson noise. In this regime, it is difficult to quantify the
properties of the stream, and it is therefore hard to determine
whether the lack of variation of the velocity dispersion at late times
is due to numerical limitations or is a real effect.

To gain insight into these issues, we will analyse the expected
behaviour of streams evolving in a smooth time-independent potential,
which should resemble that of the galaxy halo.  Using the approach
developed by HW, we can follow the evolution of streams produced in a
spherical and static NFW potential. The basic idea here is to map the
initial system onto action-angle space, then follow the much simpler
evolution in this space, and finally transform back {\it locally} onto
observable coordinates (all these being linear transformations; for
details see HW). This method, which uses action-angle variables, is
limited to applications in which the potential is separable (cf
Goldstein 1953; Binney \& Tremaine 1987). This includes all
spherically symmetric potentials but only few axisymmetric and
triaxial ones, such as the general class of St\"ackel potentials
(e.g. Lynden-Bell 1962; de Zeeuw 1985; Dejonghe \& de Zeeuw 1988).

We also approximate the phase-space density around the reference
particle by a multivariate Gaussian distribution. This is possible
because the multivariate Gaussian is determined from the properties of
particles in a volume much smaller than the size of the halo.

As discussed in Section 2.2, the (galaxy) halo can be fit by an NFW
profile:
\begin{equation}
\label{eq:rho_NFW}
\rho(r) = \rho_0 \frac{\delta_c r_s^3}{r(r+r_s)^2},
\end{equation}
where $\rho_0 = 3 H^2(z)/(8\pi G)$ and $\delta_c$ is a function of the
concentration of the halo $c_{\rm NFW} = r_{200}/r_s$:
\begin{equation}
\delta_c = \frac{200}{3} \frac{c_{\rm NFW}^3}{\log(1+c_{\rm NFW}) - 
c_{\rm NFW}/(1+c_{\rm NFW})}.
\end{equation}
Recall that $H(z) = H_0 \sqrt{\Omega_\Lambda + \Omega_0 (1+z)^3}$.
The potential associated with this density can be obtained by
integrating Poisson's equation, and is found to be:
\begin{equation}
\label{eq:phi}
\Phi(r) = -\Phi_0(z) \frac{r_s}{r} \log\big(1 + \frac{r}{r_s}\big). 
\end{equation}
Here $\Phi_0(z) = 3/2 \, H^2(z) \delta_c r_s$.  The values of the
parameters $r_s$ and $r_{200}$ given in Section 2.2 should be
multiplied by the factor $\gamma$ for the scaled halo.

In this (spherically averaged) potential we integrate the orbit of the
reference particle in time. This integration is done in two
complementary ways:
\begin{enumerate}
\item starting from the position and velocity of the reference
particle at the ``time of formation of the stream'' (this is the time
when the particles become unbound from their parent halo). In this
case, the integration is performed forwards in time, and the potential
used is given by Eq.~(\ref{eq:phi}), where $z$ is the redshift of
formation of the stream.

\item starting from the position and velocity of the reference
particle at the present time. The orbit is then integrated backwards
in time, until the ``time of formation of the stream'', in the
present-day potential ($z=0$). 
\end{enumerate}
In both cases, the orbits are integrated in the reference frame of the
centre of mass of the main progenitor of the galaxy. 

\begin{figure}
\psfig{figure=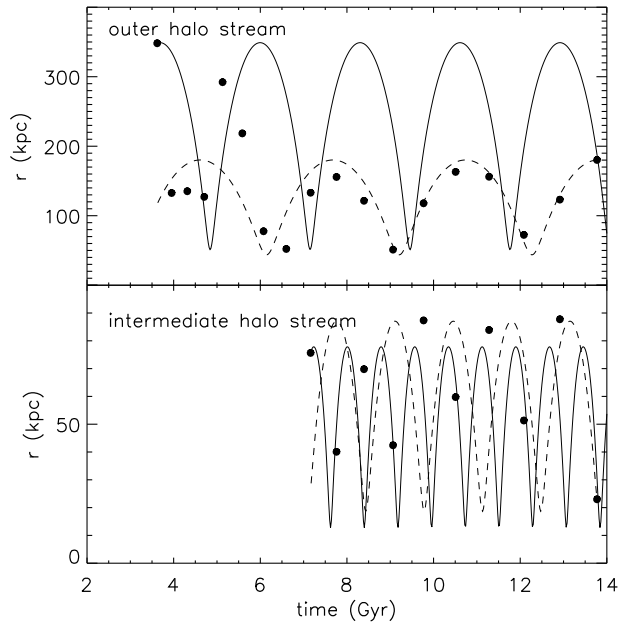,height=8.5cm}
\caption{The top panel shows the distance $r$ from the centre of mass
of the galaxy as a function of time $t$ for the reference particle in
the outer halo stream (solid circles). The dashed line corresponds to
$r(t)$ for an orbit integrated in the present-day NFW potential of the
galaxy, starting from the present-day position of this particle. The
solid line corresponds to an orbit in an NFW potential determined at
the redshift of formation of the stream (according to
Eq.\ref{eq:phi}). The bottom panel shows the corresponding $r(t)$ from
the simulation and the proposed orbits for the reference particle
moving in the intermediate regions of the galaxy halo at the present
time.}
\label{lcdm:fig_orbits}
\end{figure}

In the top panel of Figure~\ref{lcdm:fig_orbits} we plot the radial
oscillations of the orbit for the outer halo reference particle. The
solid curve corresponds to the NFW potential at redshift $z_{\rm form}
=1.8$, while the dashed curve to that at $z=0$. We note that for the
reference particle in the outer halo stream, the best agreement is
obtained when the integration is performed backwards in time.  We also
note that the potential seems to have fluctuated dramatically
until $t \sim 6$ Gyr, inducing strong changes in the radial
oscillations of the reference particle. As a whole, the orbit has
evolved to a more bound state, due to the aggregation of mass during
mergers. 

The analogous plots for the reference particle orbiting the
intermediate halo are shown in the bottom panel. Recall that here
$z_{\rm form} = 0.73$.  In this case, none of the proposed orbits fit
the actual orbit very well; it may be considered to lie in between
these two cases. Thus, the characteristics and the evolution of the
debris streams predicted for the two proposed orbits may perhaps
encompass the actual behaviour of the streams in our simulations. In
this ``intermediate halo'' case, we do not find a clear indication of
evolution in the orbit. This is likely due to the mild increase in the
mass within the orbit since $z_{\rm form}$ (see Sec.~4)

\begin{figure}
\psfig{figure=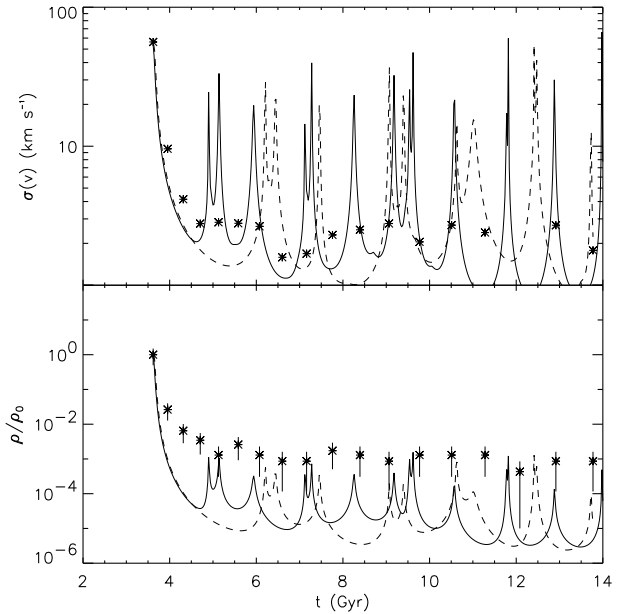,height=8.5cm}
\caption{The phase-space behaviour predicted for the outer halo debris
stream in a static NFW potential.  The top panel shows the time
evolution of the 1-d velocity dispersion, and the bottom panel the
behaviour of the density for this stream.  In both plots, the dashed
curve has been computed for the present-day NFW potential, while the
solid curve corresponds to the evolution in the NFW potential of the
time of formation of the stream. The observed properties of the stream
in the simulation are shown as asterisks and were determined from all
particles within a box of side 5 kpc centred on the reference
particle.}
\label{fig:ev_o}
\end{figure}

\begin{figure}
\psfig{figure=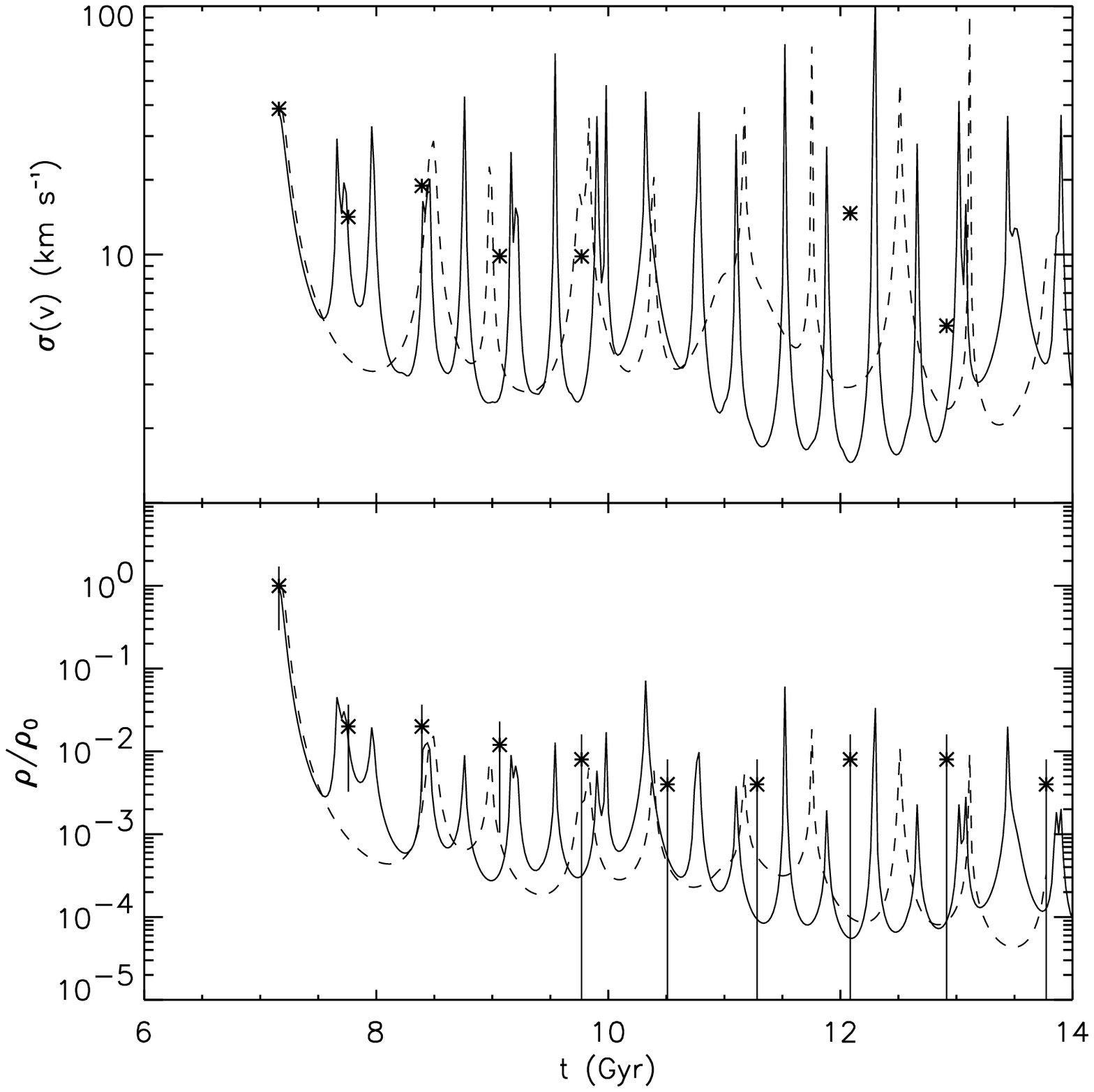,height=8.5cm}
\caption{The phase-space behaviour predicted for the intermediate halo
debris stream.  The top panel shows the time evolution of the 1-d
velocity dispersion, and the bottom panel the behaviour of the density
for this stream.  The solid curve has been computed using the NFW
potential at $z=0.73$, while the dashed curve corresponds to that at
the present-day. The observed properties of the stream in the
simulation are shown as asterisks and were determined from all
particles within a box of side 5 kpc centred on the reference
particle.}
\label{fig:ev_i}
\end{figure}

 In Figures~\ref{fig:ev_o} and \ref{fig:ev_i} we show the predicted
evolution of the 1-d velocity dispersion and the behaviour of the
density in the spherical static NFW potential for the outer and
intermediate halo streams, respectively. The actual behaviour of the
streams in the simulation is also shown for direct comparison. In
both cases, we find that the decrease observed in the simulated
streams is close to what is predicted for the evolution of streams
moving in a static potential. This is the case at least for the first
few Gyrs of evolution. Although after this period we expect some
differences, due to the fact that the orbit evolves in shape, the
agreement still appears to be quite good. We predict a rapidly varying
velocity dispersion (and density) on top of a secular evolution. The
spikes take place when particles in a stream go through a caustic
surface defined by their orbital turning points. To observe this
behaviour in the simulation would require output times spaced by 1/2
of the radial period at most, or roughly 0.25 Gyr for the stream
orbiting the intermediate regions of the halo. Our outputs, on the
other hand, are logarithmically spaced in the cosmological expansion
factor $a(t)$, so that at $z=1$ the time elapsed between two outputs
is $\Delta t = 0.52$ Gyr, and by $z=0$, $\Delta t = 0.86$ Gyr.

\section{Mass growth history of the galaxy}
\label{sec:mass_growth}

Here we focus on determining the mass growth history of the halo as a
function of distance from the halo centre. This is relevant in two
different ways. First, debris from satellites accreted at late times
will be generally less mixed, and could thus produce more massive
streams. Determining where to expect these streams will enable us to
understand the properties of the dark halo of our Galaxy. Secondly
such satellites will have had more time to form stars in them (prior
to their merging), thereby providing the galaxy with younger
stars. The time of merging could thus be used as an indicator of the
expected age distribution of stars in different regions of our Galaxy.

\begin{figure}
\psfig{figure=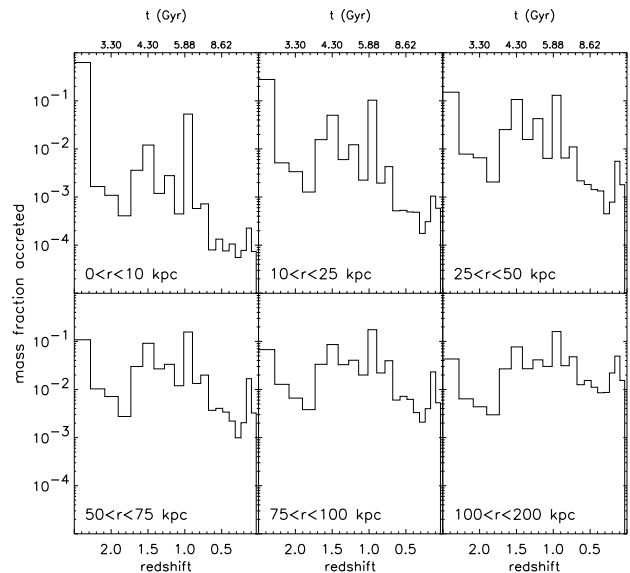,height=8cm}
\caption{Growth of mass in different spherical shells around the
galaxy centre. We notice here that a large fraction of the mass in the
inner galaxy was already in place 10 Gyr ago. For the outer galaxy
(bottom panels), the growth of mass is more gradual.}
\label{fig:shells.z}
\end{figure}

We select all satellites that merged with the galaxy since redshift
$z=2.4$, and determine when these mergers took place and what is their
final debris distribution. We proceed by dividing the halo in six
spherical shells around the galaxy centre. These shells are located
at: $ r < 10$ kpc, $10 \le r < 25$ kpc, $25 \le r < 50$ kpc, $50 \le r < 75$
kpc, $75 \le r < 100$ kpc and $100 \le r < 200$ kpc. For each particle in
a given shell, we determine when it was accreted by the main
progenitor of the galaxy.  In Figure~\ref{fig:shells.z} we show the
fraction of mass accreted normalised to the present mass for each
shell as a function of redshift (and time). We note that the formation
time of the inner galaxy is strongly biased towards high redshifts,
with more than 60\% of the mass already present at $z=2.4$ or 11 Gyr
ago.  Conversely, we find that mergers in the last 3 Gyr,
contributed a relatively small amount of mass, less than 0.1\%, to
this region of the galaxy halo. For the outer regions of the galaxy,
we note that the growth is much more gradual in time, with accretion
being almost equally important at all times.

Of course, the detailed shape of the mass growth history depends on
the detailed merger history of the halo, since the peaks observed in
the different histograms correspond to individual mergers taking place
at those times. Nonetheless, several authors have found that there is
an almost ``universal'' form of the mass accretion histories of galaxies
and galaxy clusters (e.g. van den Bosch 2002; Wechsler et al. 2002),
which would also represent reasonably well the growth of our simulated
halo. On the other hand, Zhao et al. (2002) have shown using
high-resolution simulations that the mass in the inner regions of
dark-matter halos (which they define as the mass within a scale
radius, which in our case corresponds to approximately 30 kpc), is
generally in place by $z \sim 2$. These results suggest that the
formation history of our simulated halo is rather typical. 

\begin{figure}
\psfig{figure=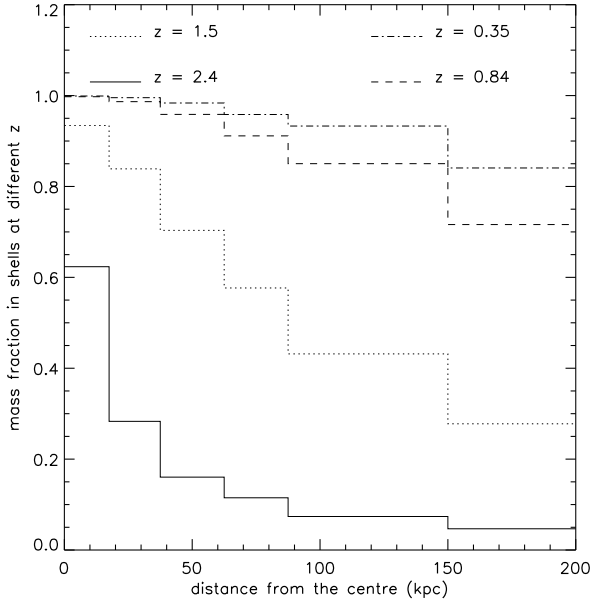,height=8.5cm}
\caption{Mass fraction present for four different redshifts as a
function of distance from the galaxy centre.}
\label{fig:shells.r}
\end{figure}

Another way of understanding the age structure of the galaxy halo is
by focusing on what fraction of the mass was in place by different
redshifts as a function of distance from the galaxy centre. To some
extent this is the cumulative distribution of the plots shown in
Fig.~\ref{fig:shells.z} for each shell. We focus on the fraction of
the present-day mass in the shell that was in place at four different
redshifts $z = 2.4$, 1.5, 0.84, and 0.35. The results are shown in
Figure~\ref{fig:shells.r}. Again we notice that 60\% of the mass in
the inner galaxy was in place by $z=2.4$, and  more than 90\% by
$z=1.5$, i.e. 9.5 Gyr ago. Thus any accreted stellar populations in the
inner regions of the galaxy are predicted to be old. On the
other hand, only about half of the particles in the intermediate regions of
the galaxy were present 9.5 Gyr ago (90\% by $z=0.84$). The formation
of the outer galaxy is more biased towards late times, with half of
its particles coming into place in the last 7 Gyr.

As a cautionary remark, let us recall that although we discuss here
the expected ``stellar" populations, we are actually describing the
growth in mass of the dark-matter halo, rather than of the stellar
halo. To describe properly the age distribution of the stellar halo we
would need to take into account that the stars in the different
satellites that merged to build up the halo probably were located
primarily in their innermost regions. This would mean, as shown in
Section \ref{lcdm_sec:phase-space}, that they should end up closer to
the galactic centre than most of the dark-matter particles of the same
satellite (see also White \& Springel 2000). Moreover, we are also
ignoring the fact that satellites orbiting the intermediate and outer
regions of the halo can survive until the present day. In those cases
it is possible that they continue to form stars, even while orbiting
inside the dark halo of the galaxy (like the Magellanic Clouds). This
means that the material that is stripped off from these systems could
also contain younger stars.

\section{The number of streams}

The results obtained in Section \ref{lcdm_sec:phase-space} suggest
that the number of streams in the inner galaxy, and in particular in
the vicinity of the Sun should be quite large. In this section we will
estimate this quantity, as well as other characteristics of the
streams, such as internal velocity dispersion and density.  We will
compute the number of streams inside cubes located in the inner
$([-60,60])^3$ kpc$^3$ in the scaled ``Milky Way" halo.  We partition
this space into boxes of 2 kpc on a side.

\subsection{General definitions}

Let there be $L$ separate true streams in a given box, each having
mass $\nu_k$ ($k=1,L$). Let us assume that the observed number of
particles in a stream follows a Poissonian distribution, and that the
actual count is $N_k$ particles ($k=1,L$). Then the expectation value
of $N_k$ is $\langle N_k \rangle = \nu_k$, and $\langle N_k^2 \rangle
= \nu_k + \nu_k^2$. (Here we measure mass in units of the particle mass
of the simulation.)

The mass-weighted mean mass per stream $\mu$ is
\begin{equation}
\mu = \frac{\sum_{k=1,L} \nu_k^2}{\sum_{k=1,L} \nu_k}.
\end{equation}
Therefore our estimate for $\mu$ corrected for Poisson noise will be
\begin{equation}
\hat{\mu} = \frac{\sum_{k=1,L} \big(N_k^2 - N_k\big)}{\sum_{k=1,L} N_k}.
\label{eq:mass}
\end{equation}
In the limit of very massive streams, the Poisson correction will be
negligible since $N_k^2 \gg N_k$. In the limit of small number of
particles per stream, the correction will be of the same order as the
quantity we measure. Note as well, that streams which in this
realization do not have any particle or just have one, do not
contribute to the numerator of Eq.~(\ref{eq:mass}). However single
particle streams do contribute to the total number of particles in the
box. If in a box we find mostly one-particle streams, then $\hat{\mu}$
can (correctly) become very small.

We define the mass-weighted number of streams $F$ in a box as the
ratio of the total mass in the box to the mass-weighted mean mass
$\mu$ per stream in the box. Therefore $F$ is 
\begin{equation}
F = \frac{\sum_{k=1,L} \nu_k}{\mu}, 
\end{equation}
and our Poisson corrected estimate of $F$ is 
\[
\hat{F} = \frac{\sum_{k=1,L} N_k}{\hat{\mu}}, \]
or
\begin{equation}
\hat{F} = \frac{\big(\sum_{k=1,L} N_k\big)^2}{\sum_{k=1,L} \big(N_k^2
- N_k\big)}.
\end{equation}
For example, if in our realization all of the streams have only two
particles $N_k = 2$, our estimate of the mass-weighted filling factor
becomes $\hat{F} \sim N_T$, where $N_T$ is the total number of
particles in the box.  In the regime where one massive stream
dominates the distribution, $\hat{F} \sim N_T/(N_T-1)$ and will thus
be close to unity. Also note that $\hat{F}$ can be larger than $N_T$ which
will happen when most of the streams contain only 1 particle.

\subsection{The minimum number of streams: the number of halos}

We would like to obtain an estimate of the number of disrupted halos
contributing to the density at each location in the galaxy. This is a
lower limit to the total number of streams present, since a halo can
(and usually does) give rise to multiple and spatially overlapping
structures, as shown in Sec.~\ref{lcdm_sec:phase-space}. This lower
limit will be particularly unrepresentative of the true number of
streams in the inner halo both because of the very short dynamical
timescales there and because more than 60\% of the mass in this region
comes from just one object identified at $z=2.4$, as shown in
Sec.\ref{sec:mass_growth}.

\begin{figure}
\psfig{figure=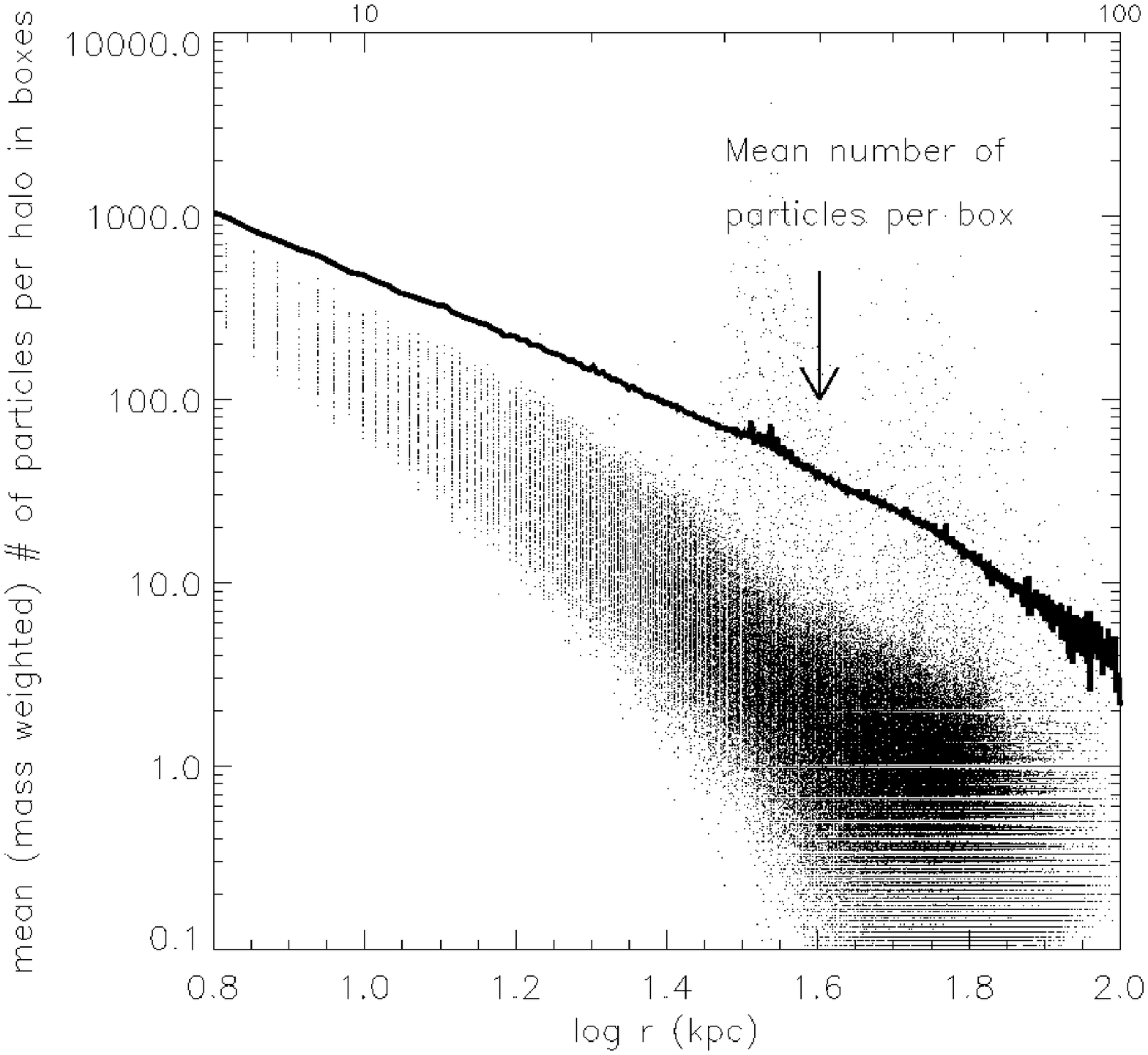,height=7.7cm}
\caption{The mass-weighted mean number of particles from halos
identified at $z=2.4$, in boxes of side 2 kpc as a function of
distance from the centre of the galaxy halo. The black curve shows the
mean number of particles in these boxes averaged over all boxes at the
same distance from the halo centre.}
\label{fig:halo_np}
\psfig{figure=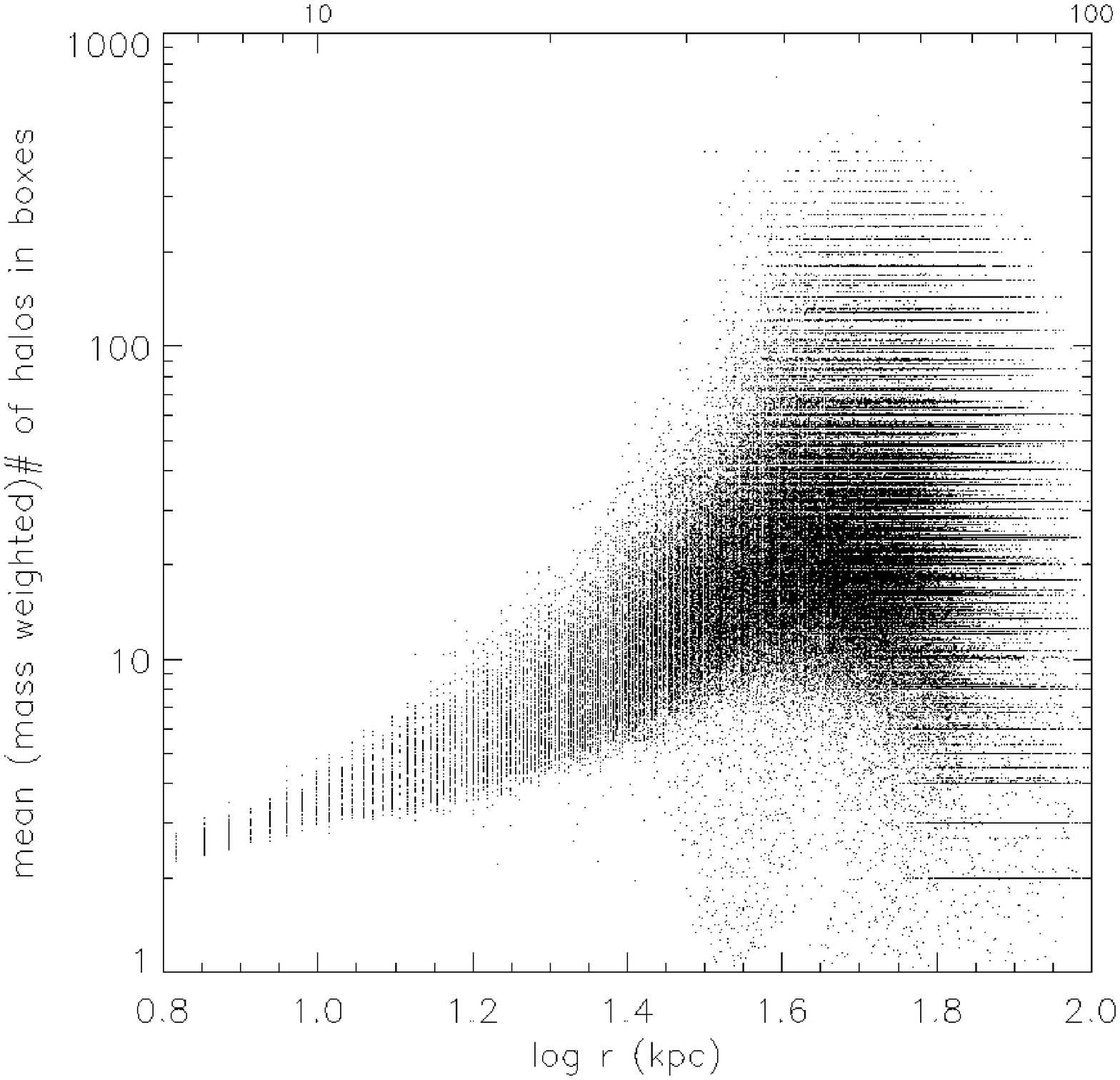,height=8cm}
\caption{The mass-weighted number of halos in boxes of 2 kpc on a side
as a function of distance from the centre of the galaxy halo.}
\label{fig:halo_ff}
\end{figure}

To obtain an estimate of the number of halos contributing to any given
position in the Galaxy, we determine which of the halos identified at
$z=2.4$ contribute to each given box and with how many particles.  In
Figure~\ref{fig:halo_np} we plot the (Poisson corrected) mass-weighted
mean number of particles per halo $\hat{\mu}_{\rm halo}$ as a function
of distance from the galaxy centre for each of the boxes
considered. The thick black line corresponds to the mean number of
particles in a box, averaged over all boxes at the same location. This
Figure shows that most of the particles in the inner galaxy come only
from a handful of disrupted halos. This can also be seen from
Figure~\ref{fig:halo_ff}, where we plot the Poisson corrected estimate
of the mass-weighted number of halos $\hat{F}_{\rm halo}$ as a
function of distance from the galaxy centre and for each one of the
boxes considered. We note that the mean number of halos per box in the
outer galaxy is large, but that each contributes only a handful of
particles. However, the inner galaxy is dominated by just a few halos
making up most of the mass.  This trend (of increasing number of halos
with distance) is due to a form of mass segregation: heavy halos can
sink by dynamical friction in short timescales to the centre of the
newly formed system, whereas lighter halos, unable to decay quickly,
can only deposit their mass in the outskirts.

\subsection{The observed number of streams in the simulations}

In general, a stream may be defined by those particles having the same
orbital phase and coming from the same halo at some initial (high)
redshift. The orbital phase of a particle may be determined, in
principle, by counting how many times it has crossed the planes $x =
0, y=0, z=0$.  However, because of the limited number of outputs which
are logarithmically spaced in the expansion factor, it is difficult to
carry this through correctly. As in section
\ref{lcdm_sec:phase-space}, we therefore define a stream as a set of
particles which
\begin{enumerate}
\item have been neighbours in all previous outputs;
\item are relatively close in space at the present time.
\end{enumerate}

\begin{figure}
\psfig{figure=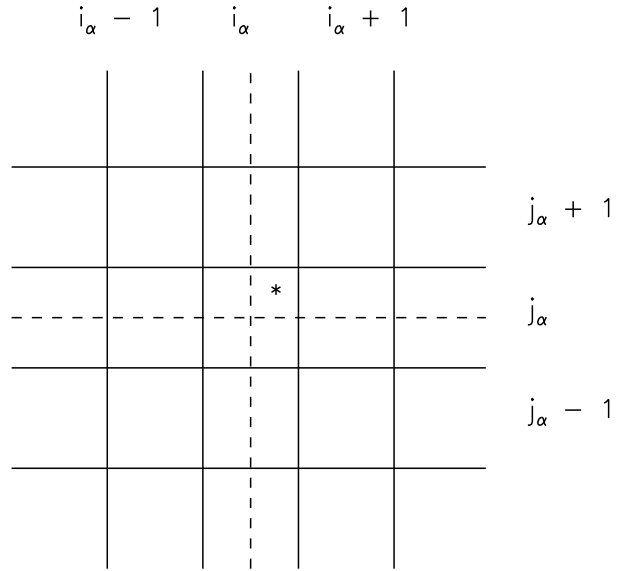,height=7.5cm}
\caption{Here we show how we tag each particle with six different
numbers, which are eventually used to determine which particles belong
to the same streams. Let the asterisk indicate the position of 
particle $\alpha$ at a certain output. The box in which this particle is
located has ID$=(i_\alpha,j_\alpha, k_\alpha)$. The closest boxes to
the particle have $i_{\alpha} + 1$, and $j_{\alpha}+1$, and for example
$k_{\alpha} - 1$ in the third dimension. For another particle $\beta$ to be
a neighbour of the particle under analysis, it ought to have
$i_\beta = i_\alpha$ or $i_\beta = i_\alpha + 1$, 
$j_\beta = j_\alpha$ or $j_\beta = j_\alpha + 1$, and
$k_\beta = k_\alpha$ or $k_\beta = k_\alpha - 1$.}
\label{fig:box_ID}
\end{figure}

In practice, we make a coarse partition of the 3-dimensional space,
whose elements are boxes of 15 kpc on a side. At each output, we check
in which box of the partition any given particle is located. We tag
the particle by this box ID, and by the IDs of those nearest to it.
So for example, a given particle will generally not be located right
at the centre of a box, but will be closer to one of its edges. Thus,
this particle is assigned six different numbers (corresponding to the
IDs of eight neighbouring boxes) for every output, as shown in Figure
\ref{fig:box_ID} for the two-dimensional analogue.  We repeat this
procedure for each output of our simulation.  Particles which are on
the same stream should have the same box tags or have been in
neighbouring boxes at all previous outputs. This is almost equivalent
to defining a box of 15 kpc on a side around each particle at each
output, and finding which particles fall within that region. Our
procedure is however, much more efficient computationally.

For the output corresponding to the present day, and for each particle
in one of the 2~kpc boxes of the partition of the
dark-matter halo, we find which other particles in that same box,
satisfy condition (i) above (condition (ii) is automatically
fulfilled).  Thus at the end we obtain a link list, where particles
have the same tags if they belong to the same stream.

\begin{figure}
\psfig{figure=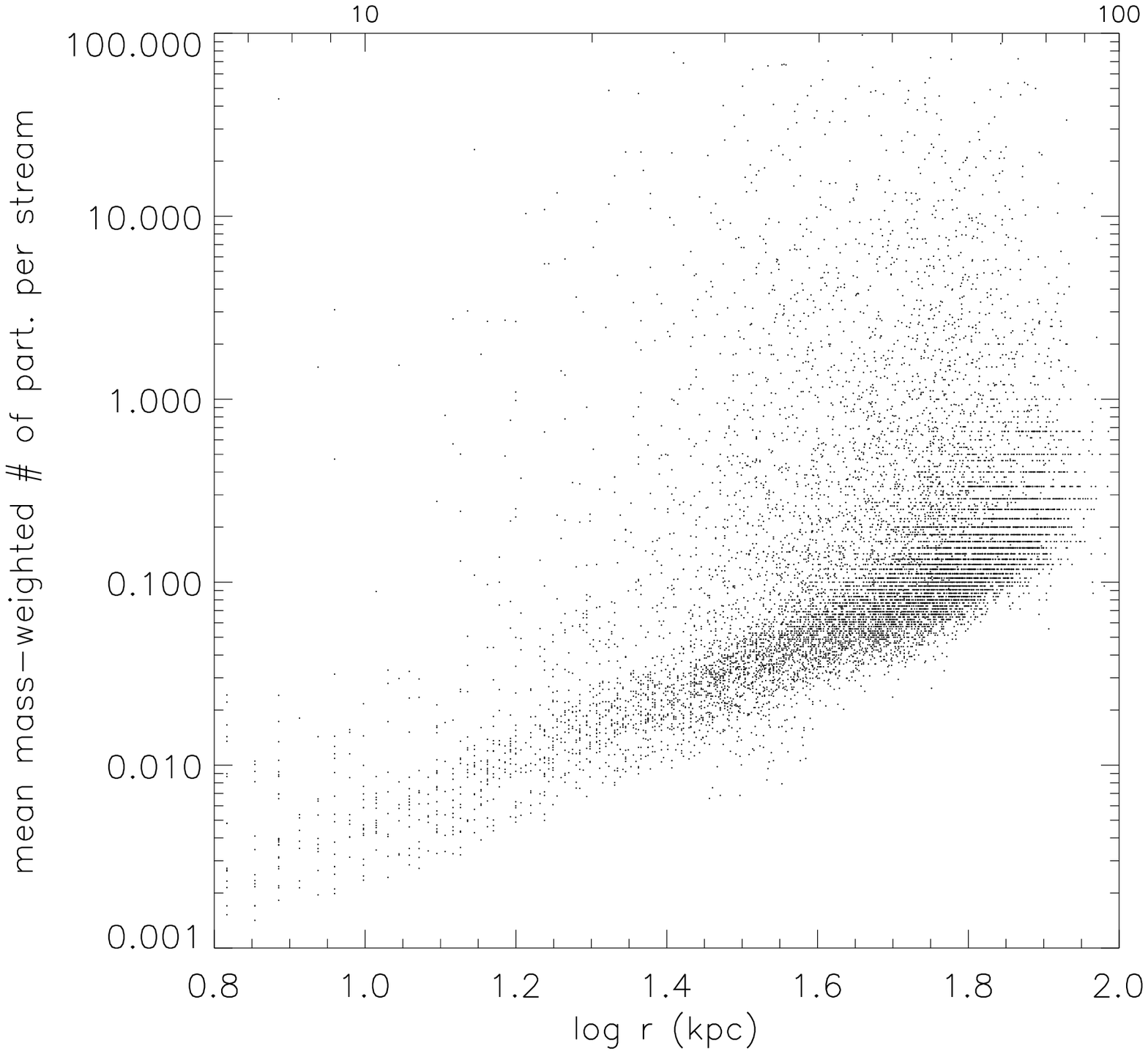,height=8cm}
\caption{The mass-weighted mean number of particles for streams in
boxes as a function of distance from the centre of the galaxy
halo. The boxes are 2 kpc on a side.}
\label{fig:np}
\psfig{figure=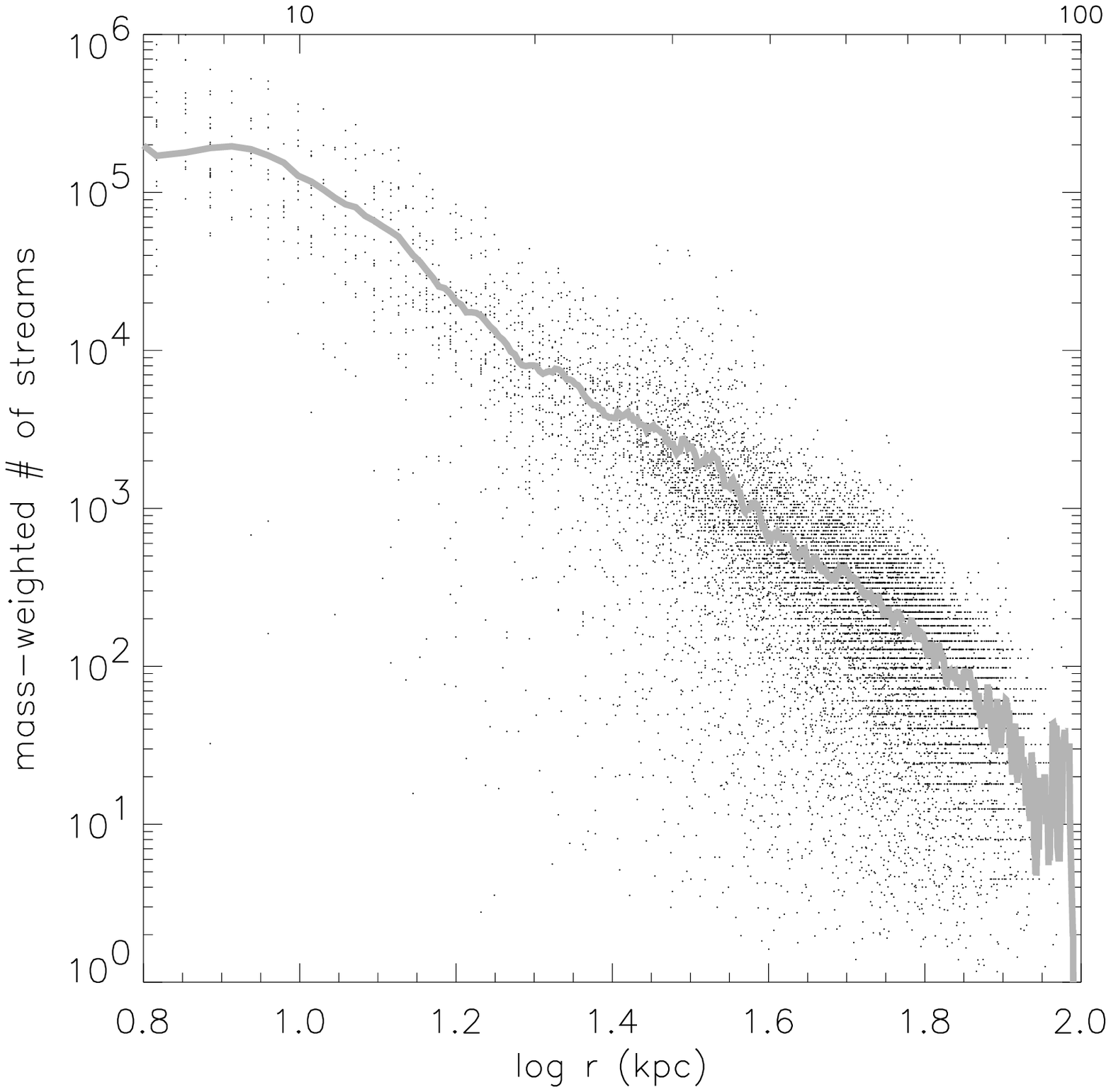,height=8.cm}
\caption{The mass-weighted number of streams in boxes of 2 kpc on a
side for our simulation S4 as a function of distance from the centre
of the galaxy halo. The dark thick line corresponds to the median
mass-weighted number of streams derived for boxes at the same distance
from the galaxy centre.}
\label{fig:ff}
\end{figure}

In Figure~\ref{fig:np} we plot the (Poisson corrected) mass-weighted
mean mass per stream $\hat{\mu}$ as a function of
distance from the galaxy centre for each one of the boxes
considered. This Figure shows that $\hat{\mu}$ is generally smaller
than unity. This happens when inside a box there are one or two
streams with several particles, and the rest are one-particle
structures. Clearly this is the case for the vast majority of the
boxes. However, at larger distances from the Galactic centre, massive
streams can be found.

In Figure~\ref{fig:ff} we plot the Poisson corrected estimate of the
mass-weighted number of streams $\hat{F}$ as a function of distance
from the galaxy centre and for each one of the boxes considered.  The
thick grey line shows the median value of $\hat{F} $, where we
consider all boxes located at the same distance from the centre.  For
the ``Solar neighbourhood" the estimate of the mass-weighted number of
streams is roughly $2 \times 10^5$ streams, showing that the local
dark halo of the galaxy is extremely well-mixed. The variations in the
number of streams at fixed distance from the galaxy centre are due to
variations in the number of particles themselves. These result from
the flattened shape of the galaxy halo.

\subsection{An analytic estimate for the number of streams}

It may at first seem counterintuitive to find a number of streams
larger than the number of particles actually observed in a box. This
effect is due to the small number of particles populating each stream,
which boosts up the Poisson correction. It is fair to say that the
number of streams we are determining is based on the actual presence
of a few hundred streams, each detected with only two or three
particles, in the inner 30 kpc region of the halo.

To ensure that the determination is meaningful we estimate how many
streams we expect to find using the analytic prescription for the
evolution of streams developed by HW. In
Sec.\ref{lcdm_sec:phase-space} we saw that this prescription does
represent a reasonable approximation to the evolution of debris
streams, even during the hierarchical build up of a CDM halo, such as
that studied here.

To obtain an analytic estimate for the number of streams expected in a
given box we proceed as follows:
\begin{enumerate}
\item We select 20 representative boxes at different distances from
the halo centre.
\item We identify the progenitor halos of the individual particles in
each of these boxes.
\item Among the particles belonging to the same progenitor halo, we
select, when possible, 10 particles (for boxes out to $r \sim 30$
kpc). In all other cases, we choose 2 particles, or just 1, if this
is all the halo contributes. 
\item For each of these particles we integrate the orbit:
\begin{itemize}
\item forwards in time: starting from the position and velocity of the
particle at $z=2.4$, using the NFW spherical potential of
Eq.\ref{eq:phi} with $\Phi_0(z=2.4)$;
\item backwards in time, until $z=2.4$, starting from the current 
position and velocity of the particle. In this case we use the NFW
spherical potential approximation to the present-day halo.
\end{itemize}
\item Once the orbit is known we can compute the evolution (always
forwards in time) of the density in the stream where the particle is
located. This is done using the action-angle formalism of HW.
\item From the median (in time) density of a stream\footnote{As
discussed in Sec.\ref{sec:mixing} the density of a stream oscillates
strongly in time. We choose to take the median density over all its
possible values in the time interval since $z=2.4$, to avoid being
dominated by a peak value just at the present time.}  we determine the
number of streams from a given progenitor halo. The number of streams
is given by the ratio of the mean (coarse-grained) density of the
debris to the actual density of the stream where the particle $i$ is
located, i.e.
\begin{equation}
N^{\rm halo}_{{\rm stream},i} = \frac{M^{\rm halo}}{V_{\rm
orb}}\frac{1}{\rho_{{\rm stream}, i}(t)}.
\end{equation}
Here $M^{\rm halo} = 4\pi/3 r_{200}^3 \langle \rho^{\rm halo}
\rangle$, $ V_{\rm orb} \sim 4\pi/3 r_{{\rm apo},i}^3$, and
$\rho_{{\rm stream},i}(t) = \rho_0 f_i(t)$, where we assume $\rho_0
\sim \langle \rho^{\rm halo}\rangle $. Therefore we can express the
number of streams as
\begin{equation}
N^{\rm halo}_{{\rm stream},i} = \left(\frac{r^{\rm
halo}_{200}}{r_{{\rm apo},i}}\right)^3 \frac{1}{f_i(t)}.
\end{equation}
\item Having obtained the number of streams for each particle, we
derive the average number of streams for each halo from all $n_p^{\rm
halo}$ particles that fall within the same box and belonged to this
same halo:
\begin{equation}
N^{\rm halo}_{\rm stream} = \frac{1}{n_{\rm p}^{\rm
halo}}\sum_{i=1}^{n_{\rm p}^{\rm halo}} N^{\rm halo}_{{\rm stream},i}.
\end{equation}
\item The total number-weighted number of streams in a box is now
obtained by adding over all halos contributing to the box, where the
weights are given by the number of particles from each halo in the box
$n_{\rm p}^{\rm halo}$:
\begin{equation}
N_{\rm stream} = \sum_{\rm halos~in~box} \big( n_{\rm
p}^{\rm halo} N^{\rm halo}_{\rm stream} \big)  \bigg / \sum_{\rm
halos~in~box} n_{\rm p}^{\rm halo}.
\end{equation}
\end{enumerate}

\begin{figure}
\psfig{figure=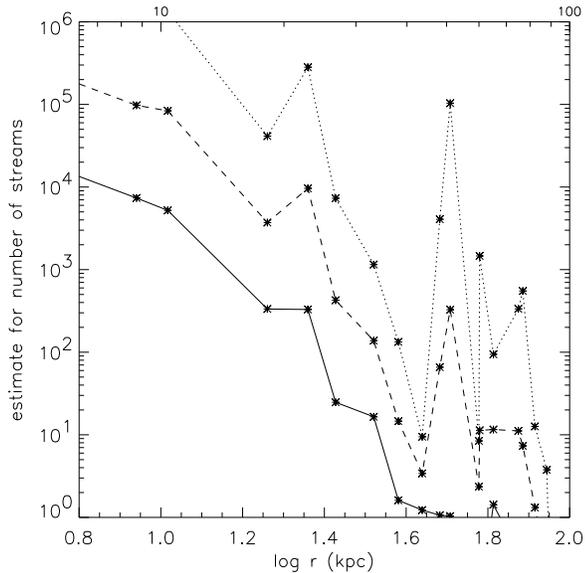,height=8.cm}
\caption{The estimated number of streams as a function of distance
from the centre of the galaxy halo. This is derived using an
action-angle formalism developed by HW, which allows one to determine
the evolution of the density of a stream as a function of time in a
given potential. The solid and dotted lines are computed for NFW
spherical potential approximations to the properties of the halo at
the present-day and at $z=2.4$, respectively. The initial conditions
of the particles' orbits are given either by their present-day
locations or by their locations at $z=2.4$.  The dashed curve is our
best estimate for the number of streams, and is the geometric mean of
the solid and dotted curves.}
\label{fig:AA_ff}
\end{figure}

In Figure~\ref{fig:AA_ff} we plot our estimate for the number of
streams as a function of distance from the Galactic centre. The solid
curve corresponds to the estimate obtained by integrating the orbit
backwards in time, from the present day conditions.  The dotted curve,
on the other hand, corresponds to the orbit integration performed
forwards in time from $z=2.4$. The geometric mean of these two
estimates is the dashed curve in the same figure.  The predicted
number of streams given by the geometric mean is in reasonable
agreement with what was shown in Figure~\ref{fig:ff}.

\subsection{Characteristics of the streams}

\begin{figure}
\psfig{figure=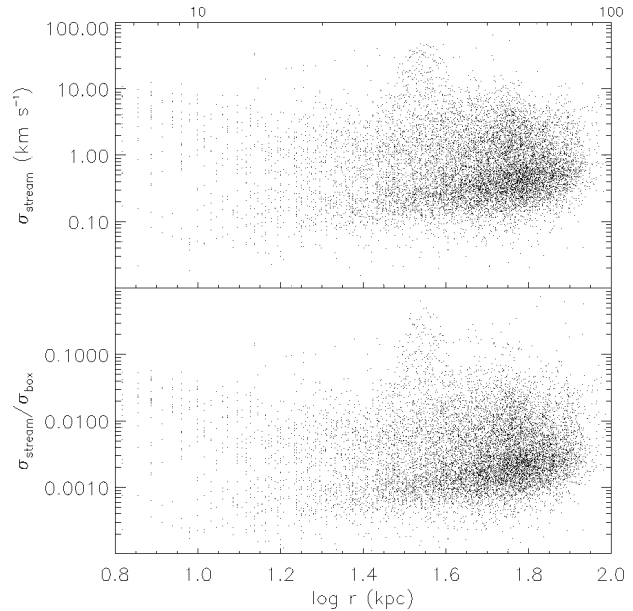,height=8.cm}
\caption{Mass-weighted velocity dispersions of streams inside cubic
volumes located in the inner ([-60,+60])$^3$ kpc$^3$ of the galaxy
halo. Boxes are 2 kpc on a side. In the bottom panel the velocities
have been normalised to the 1-d total velocity dispersion inside the
boxes at each $r$.}
\label{fig:sigma_stream}
\end{figure}

Another interesting quantity which we can calculate for each box 
is the characteristic mass-weighted dispersion within a stream:
\begin{equation}
\sigma_{\rm stream} =  \sqrt{\frac{1}{N_T} \sum_{k=1,L} 
\frac{\sum_{ij} |{\bf v}_i - {\bf v}_j|^2}{N_k - 1}}
\end{equation}
where the $ij$ sum runs over all particle pairs in the $k$-th stream
in a given box, $N_k$ is the number of particles in this stream in
this box, and as before, $N_T$ is the total number of particles in the
box. This mass-weighted velocity dispersion is expected to be
relatively small in view of our results of the phase-space evolution
of halo debris. We plot this quantity in Figure \ref{fig:sigma_stream}
as a function of distance from the galaxy centre for each one of the
boxes considered. For our boxes of 2 kpc on a side we find a typical
velocity dispersion of 1~\kms~ at the position of the ``Sun". The
values observed are probably upper limits to the typical velocity
dispersion in a stream, since they could only be measured for the
densest streams (with two or more particles), and so are biased
towards high velocity dispersions.

\section{Discussion}

We have studied the phase-space evolution of debris from the
progenitors that merge to build up a dark-matter halo in a
$\Lambda$CDM cosmology. Our analysis has shown that the debris streams
originating in progenitors of different sizes and orbital
characteristics all behave in a similar way: with velocity dispersions
and local space density decreasing in time. The evolution of the
debris streams that we were able to follow until the present time is
consistent with phase mixing. Even for halos that we could not follow
for a very long time -- because of their smaller initial number of
particles or their shorter orbital timescales-- we find the debris to
show similar behaviour. On the scales of the fine-grained distribution
function, mixing is apparently not strongly chaotic. On the contrary,
the phase-space evolution appears to be quite organised and simple,
very similar to the mixing of streams orbiting in an idealised static
potential.

In principle, a dark-matter halo formed in a $\Lambda$CDM cosmology is
not a smooth entity. Not only do dark-matter halos contain a large
number of dark satellites, they also have large amounts of
substructure in the form of streams.  We predict, however, that
dark matter in the Solar neighbourhood should be clumped in a few
hundred thousand streams, producing a velocity ellipsoid which is
close to a multivariate Gaussian (Helmi, White \& Springel
2002). These streams have their origin in the different halos that
merged to form the dark halo of the Galaxy. Most of these halos give
rise to a large number of intersecting streams in the inner Galaxy.

Determining the characteristics of these hundred thousand streams is
difficult even with the high-resolution simulation used here. We are
mostly limited by the number of particles. Although the simulation as
a whole has 66 million particles, inside a 4~kpc box centred on the
``Sun'', we find only a couple of thousand particles.  In such boxes we
find on average two hundred streams with more than one particle, and
typically each has only two particles!  The internal stream velocity
dispersion that we measure at the present-day is extremely small, of
the order of only 1 \kms.

It is encouraging that we find reasonable agreement between the
behaviour of debris streams in static potentials and that observed in
this high resolution simulation. This implies that our earlier
estimates of the number of stellar streams in the vicinity of the Sun
(Helmi \& White 1999) may indeed apply. Since the initial stellar
phase-space distribution in the progenitor objects was probably of
lower dimensionality than we here assume -- stars tend to form in
disks so that the distribution in at least two of the six phase-space
coordinates collapses -- star streams may be colder than the dark
matter streams we have analysed, and so may be more easily
distinguishable.  It is also interesting to note that because the
material that ends up populating the inner galaxy was already in place
10 Gyr ago, the oldest stars are predicted to be near the galactic
centre (White \& Springel 2000). Because this material comes from only
a few objects, one might expect the stellar populations to be quite
homogeneous, although this of course depends on whether the stars
themselves formed in these few massive objects, or whether they were
accreted into these objects in the first place.

A crude estimate of the stellar content of a stream in the Solar
neighbourhood can be obtained as follows. Let us first estimate the
``mass-to-light '' ratio per particle $[M_{\rm DM}/L_*]$ as the ratio
of dark-matter mass to stellar halo light enclosed in a shell of
thickness 2 kpc at the location of the Sun. We use the NFW profile of
Eq.(\ref{eq:rho_NFW}) for the dark-matter distribution. For the
stellar distribution we assume a density profile $\rho_*(r) =
\rho_\odot r_\odot^2 / ( r^3 + r_c^3)$, where $\rho_\odot \sim 3\times
10^4 \sm$~kpc$^{-3}$, and where $r_c$ is the core radius, and is much
smaller than $r_\odot$. We obtain for the Solar Neighbourhood $[M_{\rm
DM}/L_*] \sim 625$, for $\Upsilon_* = 2.5 \Upsilon_\odot$. For one of
our most massive streams (with 3 particles), this would imply an
average $\sim 0.54 \sm$ in stars in a sphere of 100 pc radius centred
on the Sun. Assuming a Salpeter initial mass function, and down to an
absolute magnitude of $M_V = 7^m$ (Bergbusch \& VandenBerg 1992),
i.e. $V = 12^m$ at a distance of 100 pc, this corresponds to
approximately 2 stars per stream. On the other hand, streams originate
from very localised regions of phase-space in the progenitor objects,
so it seems more likely that only about one stream in 250 has any
stars at all, that these streams have $\Upsilon_* = 2.5$
characteristic of the ``stellar'' regions of their progenitors, and that
all the other streams are dark. In this case a massive, ``luminous''
stream might contain $135 \sm$ in stars in a sphere of 100 pc radius
centred on the Sun, implying approximately $450$ stars down to $M_V =
7^m$ for a Salpeter IMF.

More reliable estimates of the stellar content and number of streams,
as well as considerably more insight into the properties of the
Galactic stellar halo would be obtained by combining semi-analytic
techniques (e.g Kauffmann et al. 1993) with high-resolution
simulations (see, for example, Springel et al. 2001).  This would
enable one to predict trends in the chemical composition, age, spatial
distribution and kinematics of halo stars as a function of position
throughout the Galaxy.

\section*{Acknowledgments}
We wish to thank the referee for the positive remarks which helped
us improve this manuscript.

\label{lastpage}
\end{document}